\documentclass[aps, prl, reprint, superscriptaddress]{revtex4-2}
\usepackage{CJK}
\usepackage{natbib}
\usepackage{graphicx}
\usepackage{bm}
\usepackage{mathrsfs}
\usepackage{hyperref}  
\usepackage{amssymb}
\usepackage{xcolor}
\hypersetup{colorlinks=true, citecolor=blue, urlcolor=blue, linkcolor=blue}

\begin{document}
\begin{CJK*}{UTF8}{gbsn}
    \title{Restricted Boltzmann machine as a probabilistic Enigma}
    \author{Bin Chen (陈斌)}
    \affiliation{State Key Laboratory of Surface Physics and Institute for Nanoelectronic Devices and Quantum Computing, Fudan University, Shanghai 200433, China}
    \affiliation{Department of Physics, Fudan University, Shanghai 200433, China}
    \author{Weichao Yu (余伟超)}
    \email{wcyu@fudan.edu.cn}
    \affiliation{State Key Laboratory of Surface Physics and Institute for Nanoelectronic Devices and Quantum Computing, Fudan University, Shanghai 200433, China}
    \affiliation{Zhangjiang Fudan International Innovation Center, Fudan University, Shanghai 201210, China}
    \date{\today}

    \begin{abstract}
 We theoretically propose a symmetric encryption scheme based on Restricted Boltzmann Machines that functions as a probabilistic Enigma device, encoding information in the marginal distributions of visible states while utilizing bias permutations as cryptographic keys. Theoretical analysis reveals significant advantages including factorial key space growth through permutation matrices, excellent diffusion properties, and computational complexity rooted in sharp P-complete problems that resist quantum attacks. Compatible with emerging probabilistic computing hardware, the scheme establishes an asymmetric computational barrier where legitimate users decrypt efficiently while adversaries face exponential costs. This framework unlocks probabilistic computers' potential for cryptographic systems, offering an emerging encryption paradigm between classical and quantum regimes for post-quantum security.
    \end{abstract}
    
    \maketitle
    \end{CJK*}
    
   \textit{Introduction}---In World War II, the mechanical cipher machine Enigma represented a milestone in cryptographic hardware implementation, using rotors and electrical circuits to perform complex substitution ciphers. However, its fixed mechanical structure and limited key space eventually led to its defeat through statistical analysis and early computing machines. Since then, cryptography has increasingly shifted towards pure software implementations. In the era of artificial intelligence, this purely digital approach faces unprecedented challenges. The inherent ability of machine learning models to exploit data patterns has raised new security concerns \cite{alani2019applications, jaschke2018unsupervised}, particularly through cryptanalysis using plaintext-ciphertext pairs \cite{andonov2020application,alani2012neuro} or data from side-channel attacks on encryption hardware \cite{hospodar2011machine}. These emerging threats underscore the necessity for robust encryption algorithms that satisfy three critical criteria: (i) Encryption algorithms must exhibit strong diffusion properties, where each bit of the ciphertext is influenced by many bits of the plaintext \cite{shannon1949communication}, ensuring that statistical patterns in the plaintext are thoroughly obscured. (ii) The encryption should rely on mathematically hard problems that are computationally intractable. For instance, the widely-used Rivest-Shamir-Adleman (RSA) asymmetric encryption scheme leverages the NP-hard factorization problem \cite{katz2007introduction, rivest1978method}, although it faces vulnerability to Shor's algorithm on quantum computers \cite{kaye2006introduction, shor1994algorithms, bernstein2017post}. (iii) The algorithm must be compatible with efficient hardware implementation. The Advanced Encryption Standard (AES), with its hardware-optimized bitwise XOR operations and strong diffusion properties, exemplifies this requirement \cite{rijmen2001advanced}. By increasing key lengths, AES can even mitigate threats from Grover's algorithm \cite{horowitz2019quantum}, which offers quadratic speedups on quantum computers. While the increasing complexity of encryption algorithms has successfully raised the computational barriers for unauthorized decryption, it has also inevitably elevated the computational costs for legitimate users. This paradox underscores the pressing need for a modern Enigma, which is a physical machine that can create an asymmetric computational barrier, enabling efficient decryption for authorized users while maintaining prohibitively high computational costs for adversaries.

In this Letter, we propose that a probabilistic computer, theoretically formulated by the model of Restricted Boltzmann Machine (RBM) can serve as a physical Enigma. The RBM, a specialized variant of the Boltzmann machine \cite{ackley1985learning}, has demonstrated capabilities in combinatorial optimization \cite{kirkpatrick1983optimization}, pattern recognition \cite{fischer2012introduction}, and as building blocks for deep belief networks \cite{hinton2006reducing}. Recent advances in probabilistic computing have enabled physical systems to function as probabilistic bits (p-bits), including memristors \cite{yan2021reconfigurable, woo2022probabilistic}, Field Programmable Gate Arrays (FPGAs) \cite{niazi2024training, patel2022logically}, magnetic tunnel junctions \cite{borders2019integer, singh2024cmos}, and manganite nanowires \cite{wang_superior_2024}. These platforms naturally implement RBM's probabilistic architecture, bridging theory and physical realization. Applications in optimization \cite{borders2019integer} and speech recognition \cite{li2024restricted} demonstrate orders of magnitude gains in efficiency \cite{bohm2022noise} compared to von Neumann architectures. As Feynman envisioned \cite{feynman_simulating_1982}, these systems efficiently simulate probabilistic phenomena, yet despite this potential, no theoretical framework exists for utilizing RBM in cryptography.

Our work aims to establish a protocol for RBM-based encryption, harnessing the natural stochasticity of these emerging hardware platforms. Specifically, we propose a symmetric encryption scheme which encodes information in RBM marginal distributions and encryption is achieved through bias permutation, which offers exponential information capacity scaling ($2^n$), factorial growth in key space, excellent diffusion properties comparable to AES, and computational complexity based on \#P problems, while simultaneously allowing efficient sampling for authorized users through specialized hardware implementations.

   \begin{figure}[t]
	\centering
	\includegraphics[width=1.2\linewidth]{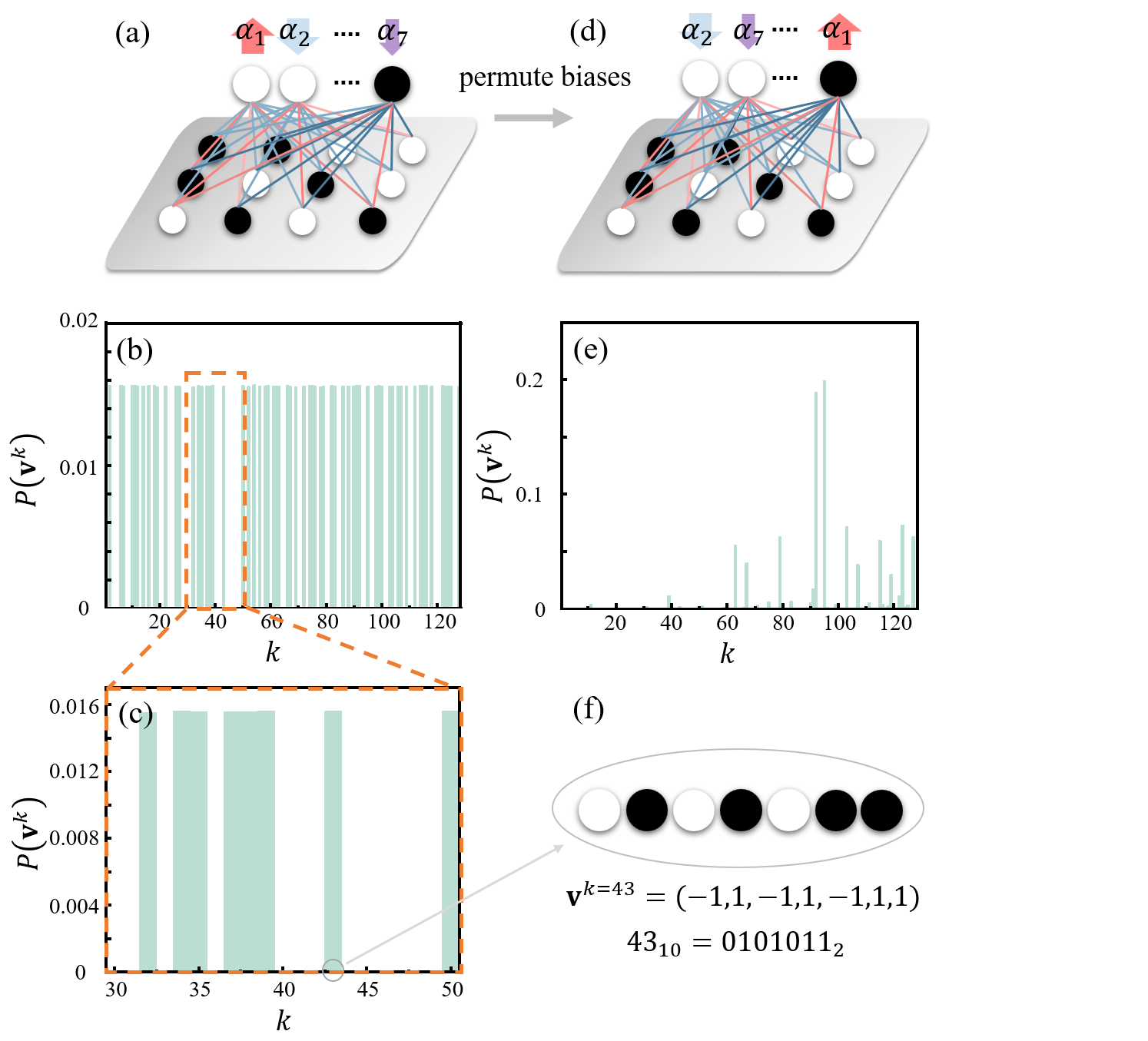}
	\caption{(a) Schematic representation of an RBM architecture consisting of a visible layer and a hidden layer with interconnecting weights. The biases applied to visible nodes are denoted by vector $\bm{\alpha}$. (b) Marginal probability distribution $P(\mathbf{v}^k)$ over the visible layer where information is encoded through training rules to determine the weights and biases. (c) Magnified view of the region highlighted by the dashed box in panel (b). (d) The identical RBM architecture with permuted visible biases, while maintaining the same weights and hidden biases. (e) Resulting marginal distribution of the visible layer after bias permutation, demonstrating significant modification on probability landscape. (f) Schematic illustration of the 43rd visible configuration.}
	\label{fig:figure1}
    \end{figure}

    \textit{Model}---As depicted in Fig.~\ref{fig:figure1}(a), an RBM is a two-layer neural network architecture comprising a hidden layer with $m$ nodes and a visible layer with $n$ nodes, where direct connections exist only between layers but not within each layer \cite{hinton2012practical}. Each node in the network represents a bipolar stochastic node that can take one of two possible states ($+1$ or $-1$) \cite{niazi2024training,hinton2012practical}.

    The equilibrium distribution of the network follows the Boltzmann distribution \cite{hinton2012practical, amin2018quantum}:
    \begin{equation}
	P(\mathbf{v}, \mathbf{h})=\frac{1}{Z}\exp(- 
        E(\mathbf{v}, \mathbf{h})),
        \label{eq:eq1}
    \end{equation}
    where the energy function $E(\mathbf{v},\mathbf{h})$ is defined as 
    \begin{equation}
		E=-\sum_{i=1}^{m}\sum_{j=1}^{n}h_{i}W_{ij}v_{j}-\sum_{j=1}^{n}\alpha_{j}v_{j}-\sum_{i=1}^{m}\beta_{i}h_{i},
        \label{eq:eq2}
    \end{equation}
    The partition function $Z$ is given by

    \begin{equation}
		Z=\sum_{k=1}P(\mathbf{v}^k, \mathbf{h}^k).
        \label{eq:eq3}
    \end{equation}

Here, $\mathbf{v}^k \in \{-1,1\}^n$ and $\mathbf{h}^k \in \{-1,1\}^m$ denote the $k$-th configuration of the visible and hidden layer nodes respectively, $W_{ij}$ represents the connection weight between the $i$-th hidden node and the $j$-th visible node, and $\bm{\alpha} \in \mathbb{R}^n$ and $\bm{\beta} \in \mathbb{R}^m$ are the bias vectors applied to the visible and hidden layers, respectively. Figure.~\ref{fig:figure1}(f) illustrates the correspondence between the index $k$ and the collective states of visible nodes, which is essentially based on decimal-to-binary conversion.

In this Letter, we propose to encode information in a \textit{distributed} manner, i.e., into the marginal (probability) distribution of the visible layer configuration $\mathbf{v}^k$ as depcited in Fig.~\ref{fig:figure1}(b) and (c), which can be obtained by summing over all possible hidden layer configurations $\mathbf{h}^k$. This marginalization can be expressed as \cite{hinton2012practical,  amin2018quantum} (see detailed derivation in Supplemental Materials (SM) \cite{SM}):
    \begin{equation}
		P\left( \mathbf{v}^k \right)=\frac{2^m}{Z}\left( \prod_{i=1}^m{\cosh}(\sum_{j=1}^{n}W_{ij}v_{j}^{k}+\beta _i) \right) \exp\left(\sum_{j=1}^{n}\alpha _jv_{j}^{k}\right).
        \label{eq:eq4}
    \end{equation}

  \begin{figure*}[t] 
	\centering
	\includegraphics[width=0.85\textwidth]{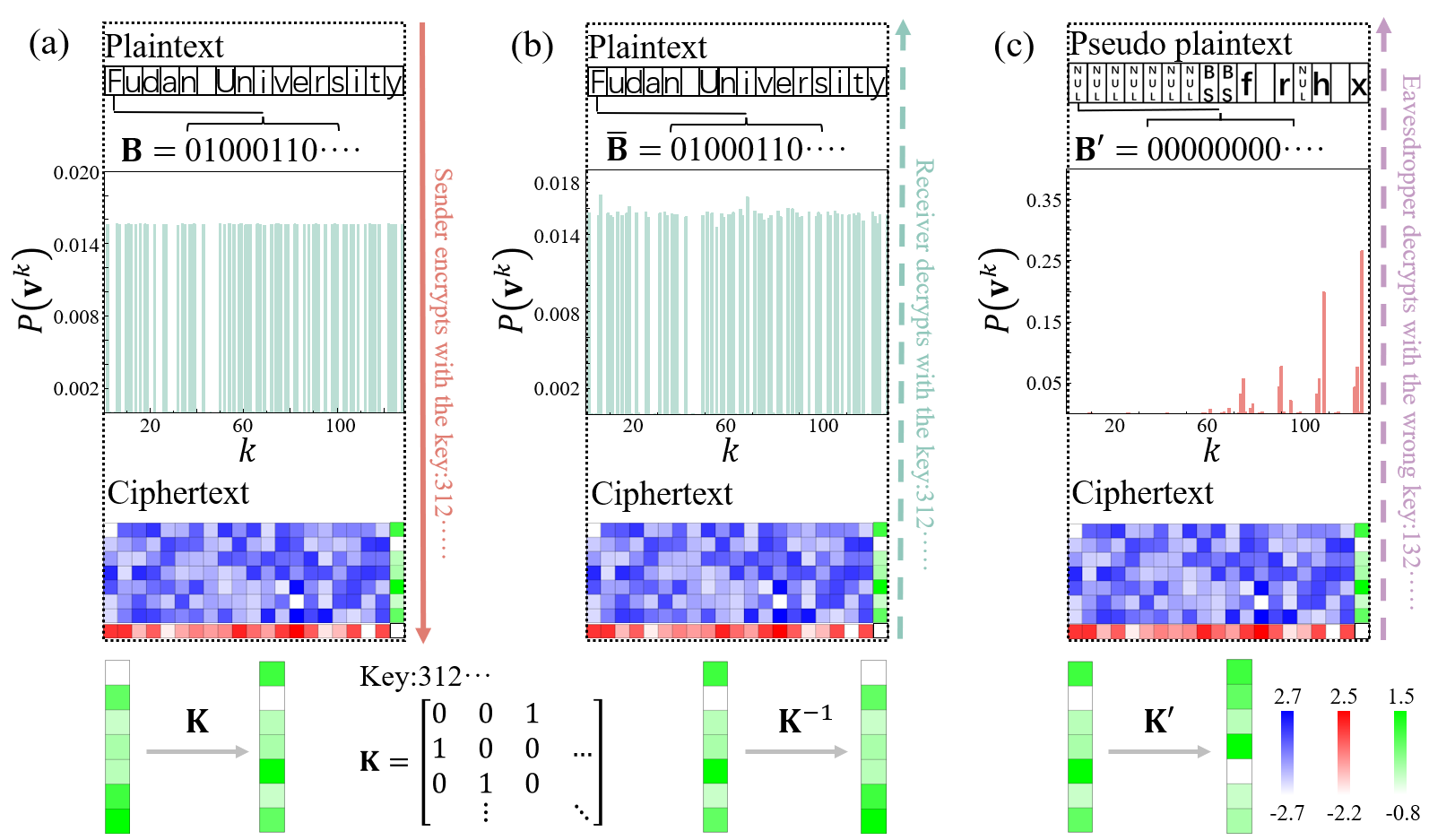} 
	\caption{(a) Encryption: The sender encodes plaintext into binary using ASCII and embeds it into the visible layer's marginal distribution. The ciphertext consists of weights (blue), hidden biases (red), and visible biases (green). A permutation matrix $\mathbf{K}$ serves as the key to permute visible biases. (b) Decryption: The legitimate receiver applies $\mathbf{K}^{-1}$ to reconstruct the RBM from the ciphertext and recovers the plaintext through probabilistic sampling. (c) Security: Without the correct key, eavesdroppers obtain only corrupted, meaningless data from sampling.}
	\label{fig:figure2}
    \end{figure*}

Given a target marginal distribution $P_t\left( \mathbf{v}^k \right)$, we aim to determine the optimal parameters set such that the marginal distribution of the RBM under parameters $\theta=\{\mathbf{W},\bm{\alpha},\bm{\beta}\}$, denoted as $P_\theta(\mathbf{v}^k)$, precisely matches $P_t\left( \mathbf{v}^k \right)$.  The optimization is achieved through a ``training'' process similar to the Contrastive Divergence (CD) algorithm \cite{ackley1985learning,hinton2012practical,le2008representational,fernandez2023disentangling,tubiana2017emergence,menezes2018handbook} with the cost function characterized by the Kullback-Leibler (KL) divergence \cite{kullback1997information, hershey2007approximating}
 \begin{equation}	                              D_\mathrm{KL}=\sum_{\mathbf{v}^k}P_t(\mathbf{v}^k)\ln\frac{P_t(\mathbf{v}^k)}{P_\theta(\mathbf{v}^k)},
\label{eq:eq5}
\end{equation}
 which is a measure that quantifies the similarity between two distributions, i.e.,  $P_t\left( \mathbf{v}^k \right)$ and $P_\theta\left( \mathbf{v}^k \right)$. We apply the gradient descent algorithm during the optimization process and the training rules are derived as $\Delta W_{\mu \rho}=-\eta \partial D_{\text{KL}}/ \partial W_{\mu \rho}$, $\Delta \alpha_{\mu}=-\eta \partial D_{\text{KL}}/\partial\alpha_{\mu}$ and $\Delta \beta_{\mu}=-\eta \partial D_{\text{KL}}/\partial \beta_{\mu}$ (see explicit expressions in SM \cite{SM}). The optimization process is set to be achieved when the KL divergence (Eq.\ref{eq:eq5}) falls below a prescribed threshold of 0.005.

Through the optimization process described in Eqs.~(\ref{eq:eq1})-(\ref{eq:eq3}), we can effectively encode information into the RBM's weights and biases $\theta=\{\mathbf{W},\bm{\alpha},\bm{\beta}\}$, establishing a mapping from parameters to the marginal distribution $P_\theta\left( \mathbf{v}^k \right)\simeq P_t\left( \mathbf{v}^k \right)$. The trained parameters provide a specific representation of the target distribution within the RBM's parameter space. This encoding ensures that any modification to these parameters, such as permutation of visible biases $\bm{\alpha}$, will result in a different probability distribution, as demonstrated in Fig.~\ref{fig:figure1}(e).

    \textit{Encryption and decryption}---  Our protocol achieves encryption by mapping a binary string converted according to the ASCII standard \cite{little1973impact} to a probability distribution across RBM visible layer configurations. As demonstrated in Fig.~\ref{fig:figure2}(a), for instance, the plaintext ``Fudan University'' is converted into a binary representation as $\mathbf{B} \in \{0,1\}^{2^n}$. This 16-character example corresponds to 128-bit binary string, requiring $n=7$ visible nodes. The target marginal probability of the $k$-th configuration $P_t\left( \mathbf{v}^k \right)$ is then determined by

    \begin{equation}
	P_t(\mathbf{v}^k)=\frac{B_k}{\sum_{i=1}^{2^n}B_i}.
    \label{eq:eq6}
    \end{equation}
    The normalization factor $1/\sum_{i=1}^{2^n}B_i$ is incorporated to ensure that $\sum_k P_t(\mathbf{v}^k)=1$. The plaintext undergoes transformation into a distinctive pattern of probability values distributed across the RBM's various configurations. Through the training process, the RBM's corresponding weights ($\mathbf{W}$) and biases ($\bm{\alpha},\bm{\beta}$) are determined. Security is established by applying an $n\times n$ permutation matrix $\mathbf{K}$ (functioning as the encryption key) to permute the visible biases, resulting in $\bm{\alpha}^\prime=\mathbf{K}\bm{\alpha}$. In this cryptographic scheme, the ciphertext consists of the weight matrix and partially permuted biases $\{\mathbf{W},\bm{\alpha}^\prime,\bm{\beta}\}$, while the permutation matrix $\mathbf{K}$ serves as the key for decryption.

    Figure \ref{fig:figure2}(b) illustrates the decryption process, where a legitimate receiver with access to both the ciphertext and key can transpose the permutation matrix $\mathbf{K}$ to obtain its inverse $\mathbf{K}^{-1}$ (note that $\mathbf{K}^T =\mathbf{K}^{-1}$ for permutation matrices). Since $\mathbf{K}^{-1}\mathbf{K}=\mathbb{I}$, the original visible biases can be restored by $\bm{\alpha}=\mathbf{K}^{-1}\bm{\alpha}^\prime$. The receiver then obtains the reconstructed plaintext using
    \begin{equation}
	\bar{B}_k=H\left(P(\mathbf{v}^k)-\frac{1}{2\times 2^n}\right),
    \label{eq:eq7}
    \end{equation}
     where the marginal distribution $P\left( \mathbf{v}^k \right)$ can be calculated directly using Eq.~(\ref{eq:eq4}), or through efficient sampling on the receiver's RBM. Here, $H$ represents the Heaviside step function. To ensure robust fault tolerance against sampling errors and maintain generality in the decryption process, we establish a criterion that a probability is classified as high if it surpasses half of $2^{-n}$, leading to $\bar{B}_k=1$, otherwise $\bar{B}_k=0$.

Figure \ref{fig:figure2}(c) depicts a scenario where an eavesdropper with access only to the ciphertext (but not the key) might attempt a brute-force attack using a random matrix $\mathbf{K}^\prime$. When this matrix is multiplied by $\bm{\alpha}^\prime$, it produces $\bm{\alpha}^{\prime\prime}$. Since $\mathbf{K}\mathbf{K}^\prime\neq\mathbb{I}$ (for $\mathbf{K}^\prime\neq\mathbf{K}^{-1}$), it follows that $\bm{\alpha}^{\prime\prime}\neq\bm{\alpha}$. The pseudo-plaintext dervied from this operation results in corrupted information that bears no meaningful resemblance to the original message. This shows that without the correct key, an eavesdropper cannot successfully recover the original plaintext.

   \begin{figure*}[t]
    \centering
    \includegraphics[width=\textwidth]{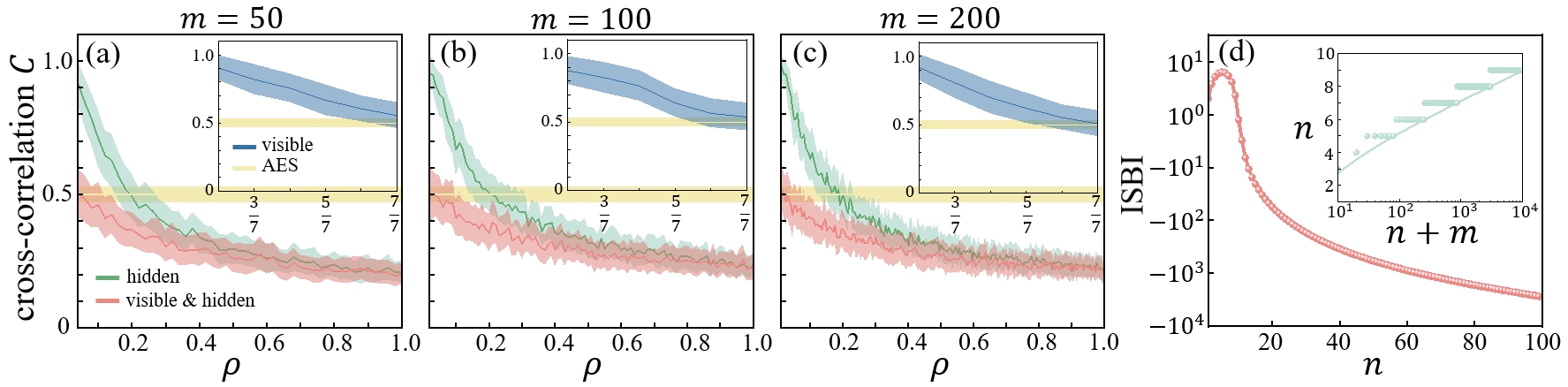}
    \caption{The dependence of cross correlation $C$ on $\rho$ (the percentage of permuted nodes relative to the total nodes) with visible nodes $n=7$ and hidden nodes (a) $m=50$, (b) $m=100$, (c) $m=200$. Mean values are represented by solid lines, and the shaded bands illustrate the variability within one standard deviation. Insets show the case where permutaion is applied only to the visible layer. (d) The dependence of the Information-Security Balance Index (ISBI) on the number of visible nodes $n$ (assuming a total node count $n+m=100$. The inset displays the number of visible nodes $n$ that maximizes the ISBI value for a given total number of nodes.}
    \label{fig:figure3}
    \end{figure*}

    \textit{Cross-correlation metric}--- The cross-correlation function serves as a quantitative metric in signal processing to measure similarity between random signals. Lower cross-correlation values indicate reduced similarity between signals. We define the cross-correlation function $C$ between true plaintext $\mathbf{B}$ and pseudo plaintext $\mathbf{B'}$ in Eq.~(\ref{eq:eq8}), 
      \begin{equation}
		C(\rho) = \frac{\mathbf{B}\cdot \mathbf{B'}(\rho)}{\mathbf{B}^2},
        \label{eq:eq8}
    \end{equation}
where $\rho$ represents the proportion of permuted nodes relative to the total nodes. We normalize $C$ by dividing by the plaintext's self-correlation product to establish a standardized measurement framework. This quantitative approach enables objective assessment of encryption quality, providing a benchmark for comparing different encryption strategies and optimizing security systems.

Figure \ref{fig:figure3}(a)-(c) show numerical simulation results of cross-correlation function $C$ versus permuting ratio $\rho$ for RBMs with $n=7$ visible nodes and $m=50/100/200$ hidden nodes encrypting ``Fudan University''. Statistical analysis reveals that $C$ consistently decreases with increasing $\rho$ across all configurations: visible-only permuting (blue), hidden-only permuting (green), or combined permuting (red). Notably, at $\rho=0.2$, all RBM configurations achieve a lower $C$ than AES (Advanced Encryption Standard, yellow), which remains at $C\approx0.5$ due to its equal bit probabilitiy in the pseudo plaintext. Our approach performs better because higher values of $\rho$ not only change the marginal distribution, but also concentrate probability mass onto fewer configurations, leaving most configurations with $P(\mathbf{v}^k)\simeq0$. An important insight is that the effectiveness depends on the proportion of nodes being permuted, not the total count of permuted nodes.

\textit{Optimization under limited resources}--- For an RBM with limited resources, i.e., total number of nodes is fixed, we investigate the optimal structural configuration to simultaneously maximize information transmission capacity and cryptographic security through key space size. Taking $m+n=100$ as an example, we analyze the corresponding key space of size $m!\times n!$. When nodes are equally allocated ($n=m=50$), the key space reaches a minimum size of $(50!)^2\approx 9\times 10^{128}$. Even at this minimum, the security remains formidable,  since the state-of-art supercomputer (El Capitan, operating at 2.746 exaFLOPS \cite{chang2024survey}) could only explore $1\times 10^{120}$ keys in $1\times 10^{95}$ years. We further define the Information-Security Balance Index (ISBI) 
\begin{equation}
\text{ISBI} = n\log_{10}\frac{mn+m+n}{2^n},
\end{equation}
 where the prefactor $n$ represents the information entropy of $n$ nodes proportional to information transmission capacity \cite{gray2011entropy}, and the logarithmic term describes the average number of tunable parameters allocated to each configuration, with larger values suggesting enhanced encoding capacity of the model. The ISBI provides a balanced metric that optimizes both information transmission and model flexibility simultaneously. Our analysis reveals that the optimal resource allocation that maximizes the ISBI occurs at an asymmetric distribution of $n=5$ visible nodes and $m=95$ hidden nodes, as indicated in Fig.~\ref{fig:figure3}(d). The maximum can be obtained by letting $\partial \text{ISBI}/\partial n=0$ (See SM \cite{SM}). As shown in the inset of Fig.~\ref{fig:figure3}(d), while the optimal allocation of $n$ grows slowly with increasing $n+m$, the amount of information that can be encoded grows exponentially as $2^n$.

   \textit{Physical implementation on probabilistic computers}--- The security of any cryptographic system fundamentally hinges on the temporal asymmetry between information validity and decryption timescales. For our RBM-based scheme, this requires the legitimate receiver's decryption speed to surpass both the message expiration time and any potential eavesdropper's computational capability. Two distinct approaches exist for recovering plaintext from the ciphertext: (i) exact computation through $P\left( \mathbf{v}^k \right)$ evaluation (Eq.~\ref{eq:eq4}) or (ii) statistical estimation via sampling.

\begin{figure}[h] 
		\centering
		\includegraphics[width=1.05\linewidth]{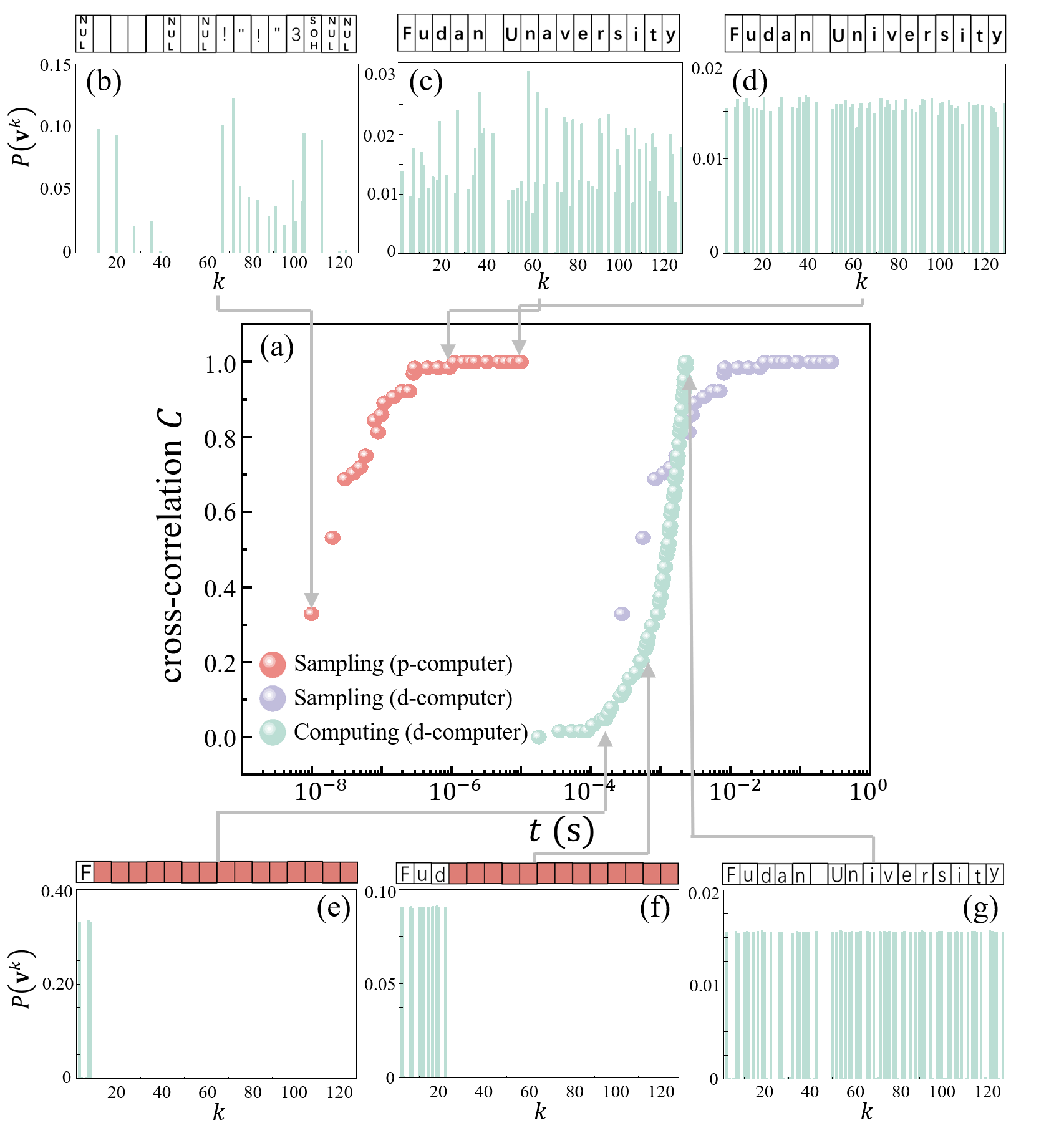} 
		\caption{(a) Decryption accuracy (characterized by cross-correlation $C$ between original text and decrypted text) as a function of decryption time for different methods: computing $P\left( \mathbf{v}^k \right)$ using digital computers (green dots), sampling using digital computers to simulate RBM (purple dots), and estimated sampling using probabilistic computers (red dots). (b)-(d) Decryption through sampling to obtain marginal distribution and corresponding plaintext with (b) 1,000 samples, (c) 100,000 samples, and (d) 1,000,000 samples. (e)-(g) Decryption by computing $P\left( \mathbf{v}^k \right)$which is computed from left to right, with plaintext characters sequentially revealed in order.}
		\label{fig:figure4}
    \end{figure}

Computing the partition function for our encryption scheme belongs to the \#P-complete class \cite{regts2018zero, valiant1979complexity, liu2021tropical}, a complexity category strictly harder than the NP problems underlying RSA factorization, ensuring that brute-force decryption via partition function evaluation scales exponentially as $ \mathcal{O}(2^n m n) $. To demonstrate practical implications, we implemented decryption for the minimal problem instance ($n=7, m=20$) on standard GHz-class digital computers. Quantitative analysis using cross-correlation metrics (Fig.~\ref{fig:figure4}(a), green dots) reveals a linear time dependence in the information recovery rate. Decryption achieves bitwise accuracy by sequentially computing $P\left( \mathbf{v}^k \right)$ from left to right, revealing plaintext characters in order, while partial outputs remain cryptographically secure.

In stark contrast, sampling-based decryption exhibits linear scaling with sample size $N$, and its statistical error diminishes as $\frac{1}{\sqrt{N}}$. Although the convergence rate gradually slows down, practical decoding typically achieves sufficient accuracy well before reaching theoretical limits (purple dots in Fig.~\ref{fig:figure4}(a) simulated by digital computers). This motivates the proposed acceleration through probabilistic computing architectures. Probabilistic computers implement natural sampling accelerators through their physical embodiment of stochastic bits (p-bits) in Ising-type systems \cite{parks2018superparamagnetic, kaiser2019subnanosecond,camsari2019scalable,borders2019integer,camsari2019p,camsari2017stochastic,grimaldi2022spintronics}. These architectures directly emulate the RBM's steady-state distribution through intrinsic thermal fluctuations, achieving sampling rates of $ 10^{11} $~samples/s \cite{chowdhury2023full} with sub-fJ/operation efficiency \cite{zand2018low,vodenicarevic2017low}. Such performance enables rapid convergence to the threshold sample count $ N > 2^{2(n+1)} $, at which point the sampling error $ 1/\sqrt{N} $ falls below the decoding criterion $(2\times 2^n)^{-1}$ in Eq.~(\ref{eq:eq7}). This ensures reliable decoding as indicated by the red dots in Fig.~\ref{fig:figure4} estimated for probabilistic computers. Since the marginal distribution depends only on $n$, the sampling method remains unaffected as $m$ increases. As a result, these architectures effectively create a probabilistic analog of the Enigma machine, significantly shortening decryption timescales. Even if an eavesdropper obtains the correct key, the information would quickly expire unless they employ a specialized probabilistic Enigma device.

Our method does not rely on integer factorization or the discrete logarithm problem, rendering it immune to Shor's algorithm. While Grover's algorithm provides a quadratic speedup for brute-force search, doubling the key length suffices to preserve the original security level \cite{kaplan2015quantum}. Moreover, large-scale, fault-tolerant quantum attacks would require millions of physical qubits, which remains infeasible in the near future \cite{bernstein2017post,campbell2017roads}. Consequently, our approach offers robust security against both current and anticipated quantum computing capabilities.

    \textit{Conclusions}--- This work presents a proposal for probabilistic Enigma, a symmetric encryption framework leveraging Restricted Boltzmann Machines and bias permutation keys. By encoding information into marginal probability distributions, our approach creates a vast key space and robust diffusion, forming a significant barrier to decryption. The design is particularly well suited to probabilistic computing hardware, allowing efficient decryption by intended users and increased computational difficulty for adversaries. Unlike conventional methods tied to hard mathematical problems, our scheme utilizes rapid fluctuations inherent in probabilistic computers or fluctuating physical systems, offering an adaptive and complementary paradigm for cryptography rather than replacing established systems like AES. Most notably, this work unlocks the practical and theoretical potential of probabilistic computing in cryptography, establishing a foundational framework that invites further exploration of its security applications.

\textit{Acknowledgements}--- This work was supported by the Innovation Program for Quantum Science and Technology (Grant No. 2024ZD0300103), National Key Research Program of China (Grant No. 2022YFA1403300) and National Natural Science Foundation of China (Grant No. 12204107). The authors thank Hangwen Guo and Jiang Xiao for inspiring discussions.

    \bibliographystyle{apsrev4-2}
%
\end{document}